# Multi-Resonant Laser Isotope Separation

Mark G. Raizen[1] and Aaron D. Barr
Dept. of Physics, University of Texas at Austin, Austin, TX 78712, USA
E-mail: raizen@physics.utexas.edu

**Abstract**

A new method for efficient isotope separation is proposed. It is based on efficient photoionization of atoms by a continuous-wave laser using resonant-enhancement in an ultra-large volume optical cavity. This method should enable higher efficiency than the existing state of the art and could be used as an alternative to radiochemistry. It should also allow separation of radioisotopes that are not amenable to standard radiochemistry, with important implications for medicine.

**Introduction**

Most elements in the periodic table have several stable isotopes differing from each other by the number of neutrons in the nucleus. There are 254 stable isotopes that are found with varying natural abundances, created in the early universe. Beyond stable isotopes, the list is greatly expanded when one includes radioisotopes of the elements, with half-lives ranging from short (less than 1 second) to intermediate (days or weeks) to long (years). These radioisotopes typically decay by emission of high-energy particles, including electrons (betas), helium nuclei (alphas), and photons (gammas) [1].

Isotopes are a great resource for humanity. In medicine, therapy with radioactive seeds is used for treating tumors [2]. Radio-immunotherapy, where cancer cells are directly targeted and destroyed, offers treatment for metastatic cancer, and holds potential for even higher efficacy as targeting molecules are improved [3]. Medical imaging with radioisotopes provides diagnosis of illness [4]. Stable isotopes can be used as tracers for the diagnosis and treatment of metabolic disorders, and for early detection of disease by a change in isotopic ratios in the body [5].

The first step towards realizing any applications of isotopes is the ability to separate them, a very difficult task due to their nearly identical physical and chemical properties. The two main methods to date are the calutron, invented over 80 years ago by Ernest Lawrence [6], and the gas centrifuge, which is only applicable to a few elements in the periodic table that have gas-phase compounds at room temperature [7]. The calutron is the most general method for isotope separation, now referred to as electromagnetic isotope separation (EMIS). While EMIS is an established method, it is energy intensive due to the large magnetic field required over the entire volume. The large-scale machines in the U.S. and elsewhere were shut down in the late 1990's due to their very high operating costs, leaving only Russia as the global source. Near total dependence on subsidized Russian machines created a fragile supply chain [8]. In recent years, the US Dept. of Energy has invested significant resources towards construction of new EMIS units to address growing demand and this effort has been successful [9].

---

[1] Author to whom any correspondence should be addressed.

Motivated by the desire to make isotope separation machines more compact and energy-efficient, the possibility of using lasers for this purpose has been the topic of intense effort. Previous work on laser separation of isotopes relied on multiphoton ionization with pulsed lasers [10]. In some variations, continuous-wave (CW) lasers were proposed and used to achieve isotopically selective optical pumping into metastable states, but the final step of laser-ionization was always accomplished with a pulsed nanosecond laser [11-14]. This method, known as Atomic Vapor Laser Isotope Separation (AVLIS), was developed for uranium separation by large scale government programs as an alternative to the gas centrifuge. Ultimately, these large-scale AVLIS programs failed and were abandoned. Proposals and demonstrations of AVLIS with various stable isotopes continue to this day, yet AVLIS has never transitioned from numerical simulations and demonstration experiments to commercial production despite many years of effort. The reasons may be due to the requirement for very high laser power needed for efficient ionization, combined with low repetition rate of the pulsed lasers. The need for continuous operation demands robust lasers and other components. Meanwhile, there has been great progress in solid-state laser technology over the past years, providing robust platforms of tunable continuous-wave lasers at the several Watt power levels. These include semiconductor diode lasers [15] and vertical external cavity surface emitting (VECSEL) lasers [16].

In recent years, a new method of laser isotope separation was developed by our group, magnetically activated and guided isotope separation, (MAGIS) [17-18]. It uses lasers to change how atoms respond to a magnetic field through a process known as magnetic-state optical pumping. Desired isotopes are then separated using arrays of permanent magnets, which act as a guiding track. The isotopically selective laser excitation relies on an isotope shift of atomic resonant frequencies that depends on nuclear mass. The lasers are low-power and solid-state, thereby very efficient and reliable. The isotope shifts are typically in the range of 1 GHz out of an optical frequency, so only one part in $10^5$, but that shift is larger than the Doppler profile of the atomic beam. MAGIS relies on permanent rare-earth magnets to create a large field very close to the surface of the magnets. It therefore does not consume much energy, unlike the calutron, except for vaporization of the element in a crucible, operation of the vacuum pumps and the lasers (taken together, these are typically around three orders of magnitude smaller than electrical consumption by the calutron). MAGIS was proven in the laboratory [19] and protected by an issued patent [20]. MAGIS is being used to produce commercial quantities of Yb-176 at very high enrichment needed for production of the radioisotope Lu-177 which is used for cancer therapy [21].

The limitations of MAGIS are primarily due to the currently available strength of permanent rare-earth magnets. First, to reach the field strengths required for separation, the atoms must be collimated in one dimension to less than 20 mrad so that they may be aimed at small, grazing angles of incidence to the magnetic array. This collimation also reduces atomic flux. Laser cooling can brighten the beam considerably, but at the cost of much higher laser power [22-23].

Second, scaling up production requires large magnetic arrays, precluding the application of MAGIS for separation of radioisotopes. The safe handling of radioisotopes, especially gamma-emitters, requires a special enclosure that provides radiation shielding for operators together with robotic control, known as a hot-cell [24].

In this paper, we propose a new method for isotope separation based on photoionization of atoms that should resolve the current limitations of AVLIS. The method, Multi-Resonant Laser Isotope Separation (MRLIS) is described in detail for a generic case, and then specifically for strontium isotope separation. Applications to several isotopes are outlined, both stable isotopes and radioisotopes as an alternative to radiochemistry.

**MRLIS description**

The goal is to efficiently ionize the desired isotopes in an atomic beam using lasers, and an electric field to separate them from the atomic beam so they can be collected. The approach used in the past for laser ionization relied on a fast sequence of high-intensity nanosecond pulsed lasers. The excitation of atoms from the ground state using resonant atomic transitions can be very efficient, but the problem was with the last step of photoionization.

We propose to use continuous-wave lasers to drive atomic transitions, combined with long interaction time in ultra-large volume resonant cavities. This new paradigm has many advantages, including the use of reliable solid-state lasers and predicted high ionization efficiency.

A schematic of the proposed set-up is shown in Fig. 1. The first step in the process is excitation of the desired isotopes from their ground electronic state to an excited electronic state. This is not fundamentally different than what is done in AVLIS, especially with recent clever proposals to use CW lasers or time-delayed pulsed lasers to pump into a metastable atomic state [11-12]. In some cases, a two-step excitation from the ground state should be employed, as shown in the figure, which can be driven efficiently using stimulated Raman adiabatic passage [25].

The last step is photoionization which is non-resonant and typically requires a high-power pulsed laser. We propose to use an optical resonator to boost the laser power by a factor of at least 1000x. The problem with optical resonators, also known as Fabry-Perot interferometers, is that the Gaussian waist scales as the product of cavity length and mirror radius of curvature to the one-fourth power [26]. The resulting beam waist is usually in the range of 100 micrometers, which would intersect only a small fraction of the atomic beam and result in very low average ionization probability. One option is to use a cavity with two flat mirrors and an intracavity lens, but the beam waist is on the order of several mm [27]. The alternative is to make the radius of curvature very large using thin-film deposition. In previous work, a radius of curvature of 10 km was demonstrated [28], but could we make it much larger? The answer is yes, simply by using flat mirrors! Such a configuration is usually called an etalon, but in the past, it has been used with low reflective coatings or none at all. Such a configuration with high-reflectivity mirrors was first considered in [29], though applications to laser ionization were not proposed.

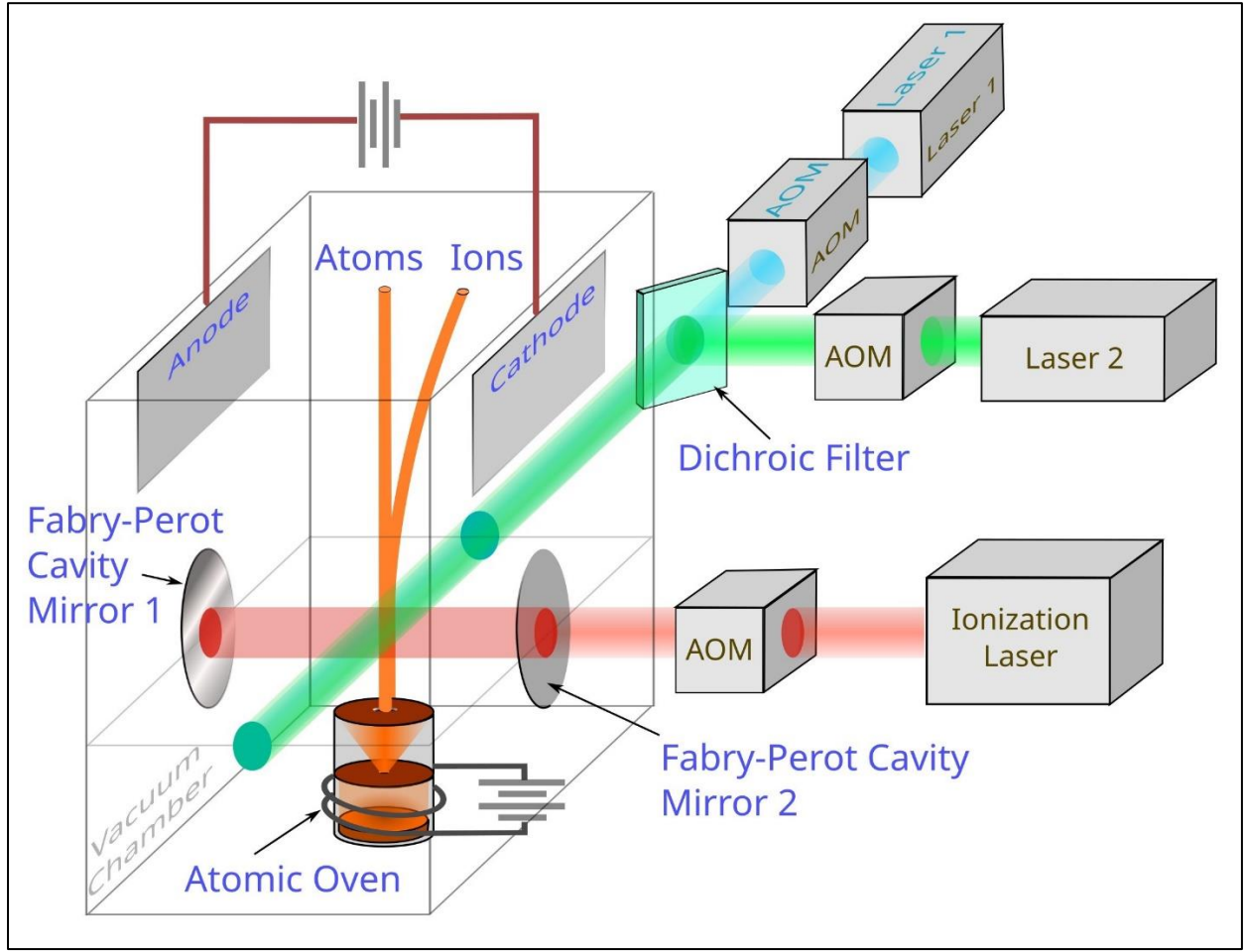

**Fig. 1.** Schematic figure of proposed experimental set-up. An atomic oven inside a vacuum chamber produces an effusive atomic beam that is collimated. The blue/green lasers are combined with a dichroic mirror/filter to excite the desired atoms to an excited atomic state. The ionizing laser (red) is enhanced in a large mode-volume cavity. Ionized atoms are collected on a cathode. The laser powers are controlled by acousto-optic modulators (AOMs).

The only condition for resonance is that there be a uniform phase front to the beam, and the overall length of the cavity be an integer multiple of half-wavelengths. This can also be seen by the resonant frequencies of transverse modes, which are shifted from the fundamental resonances by integer multiples of $\cos^{-1}(1-L/R)$ where $L$ is the cavity length and $R$ is the radius of curvature. In the limit of infinite R, these frequency shifts are zero, so mode-matching is not required. The Rayleigh length of a laser beam is the distance over which the transverse radius increases by the square root of 2 and is proportional to square of the minimum spot size divided by the wavelength. For example, a laser at a wavelength of 1 micrometer with a spot size of 6 cm has a Rayleigh length of about 3 km. If the mirror reflectivity is 99.9%, and the cavity length is 10 cm, then 1000 round-trips is only 100 m, and the beam does not expand significantly. A laser of power 10W would produce an intra-cavity power of 10 kW. For comparison, LIGO routinely has a circulating power of that magnitude or higher in their resonators [30]. To minimize absorption losses and heating of the coatings, near-infrared wavelengths longer than 800 nm are optimum.

To test the above hypothesis, we performed numerical simulations, and the results are shown in Fig. 2 and Fig. 3 below.

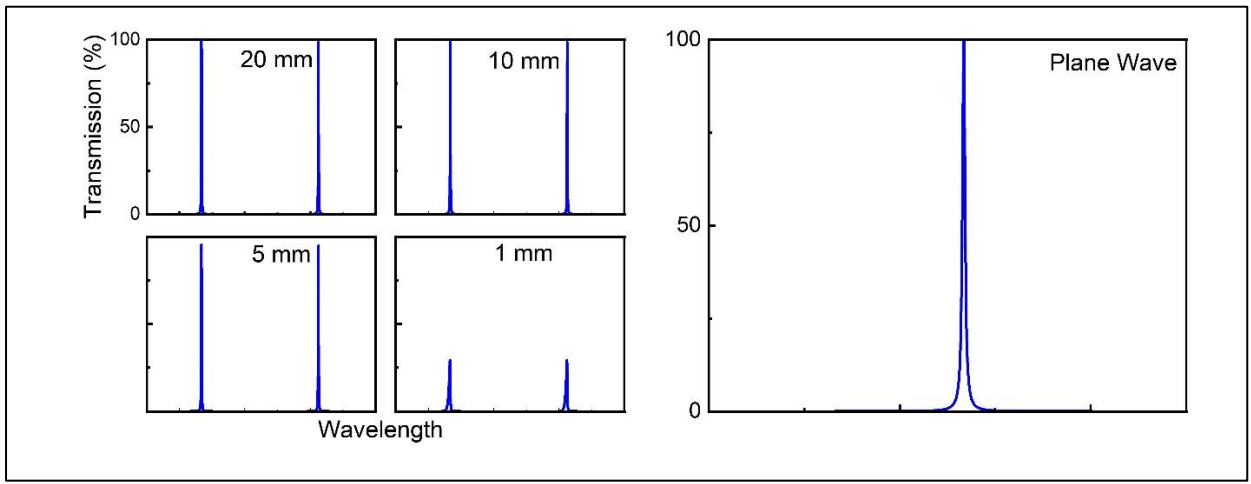

**Fig. 2.** Numerical simulation of transmission coefficient of a cavity of length 5 cm and mirror reflectivity of 99%. The spot size is changed from 20 mm to 1 mm in the four panels on the left side. For comparison, a plane wave simulation is shown on the right side.

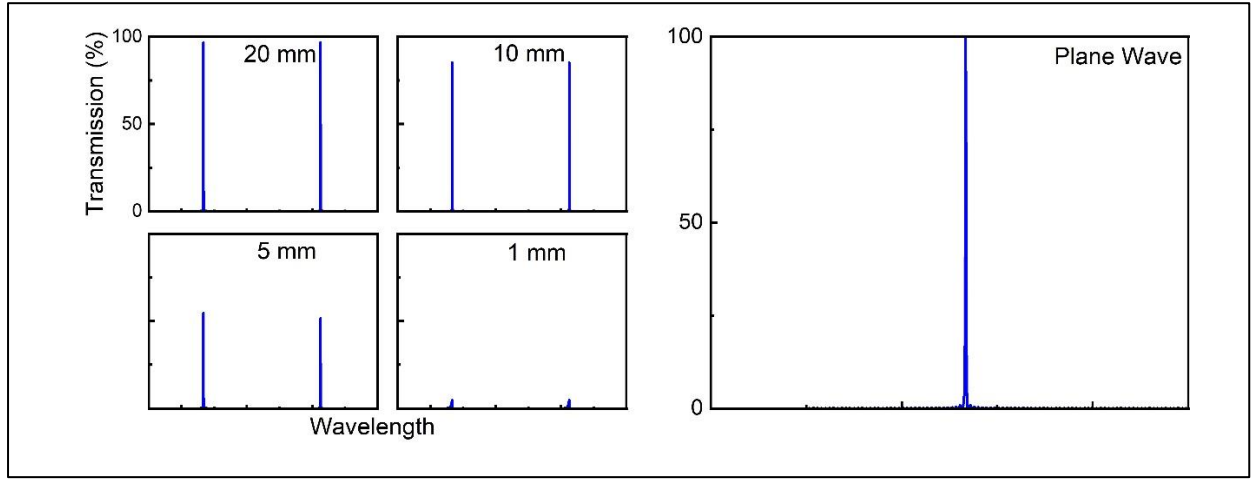

**Fig. 3.** Numerical simulation of transmission coefficient of a cavity of length 5 cm and mirror reflectivity of 99.9%. The spot size is changed from 20 mm to 1 mm in the four panels on the left side. For comparison, a plane wave simulation is shown on the right side.

The method and details of the numerical simulation are in the Appendix. The figures show the cavity transmission as the wavelength is scanned, alternatively the length of the cavity can be changed, and resonances occur for a half-wavelength change. We assume that mirror transmission T=1-R, which is justified when scatter and absorption losses are much smaller than transmission. A cavity transmission of unity is expected for a plane wave, and the intracavity power is boosted by a factor of 1/T. In the case of mirrors with 99% reflectivity, the results match the plane-wave for spot sizes above 10 mm. In the case of mirrors with 99.9% reflectivity, the spot size much be larger than 20 mm. These numerical results agree with the heuristic arguments above.

The ionizing laser inside the cavity will induce an optical Stark shift [31] on a transition being driven simultaneously with the ionizing laser due to the high intracavity power. This could be mitigated by using a magic-wavelength laser which has been employed in optical lattice clocks with great success [32]. However, if the other transition is dipole-allowed, the AC Stark shift due to the ionizing laser would not create a problem [33].

In addition to isotope separation, the above method can be used for ultra-sensitive detection of atomic isotopes. Isotope sensing is especially important for detection of rare isotopes as tracers for metabolism [34] and isotope ratios are sensitive indicators of disease [35]. Small blood or urine samples containing atomic isotopes are already analyzed using mass spectrometers, which have become extremely precise over the years. However, their cost is often prohibitive, and slow sample preparation to isolate the element of interest limits their throughput to a small number of tests per day. Laser spectroscopy can be much more sensitive than mass spectroscopy and has been demonstrated to reach laser shot-noise limited sensitivity using FM spectroscopy [36]. It is even possible to detect atoms below the standard quantum limit using squeezed light [37]. Detection of atomic isotopes using MRLIS would enable atom counting by detecting the produced ions and removing background signals by modulation of the lasers and lock-in detection of the ion current. This should allow the detection of atomic isotopes limited only by quantum projection noise, which is the ultimate limit [38]. This would be realized in an atomic beam and does not require a closed cycling transition which is needed for detection of trapped ions or with the atom trap trace analysis method [39].

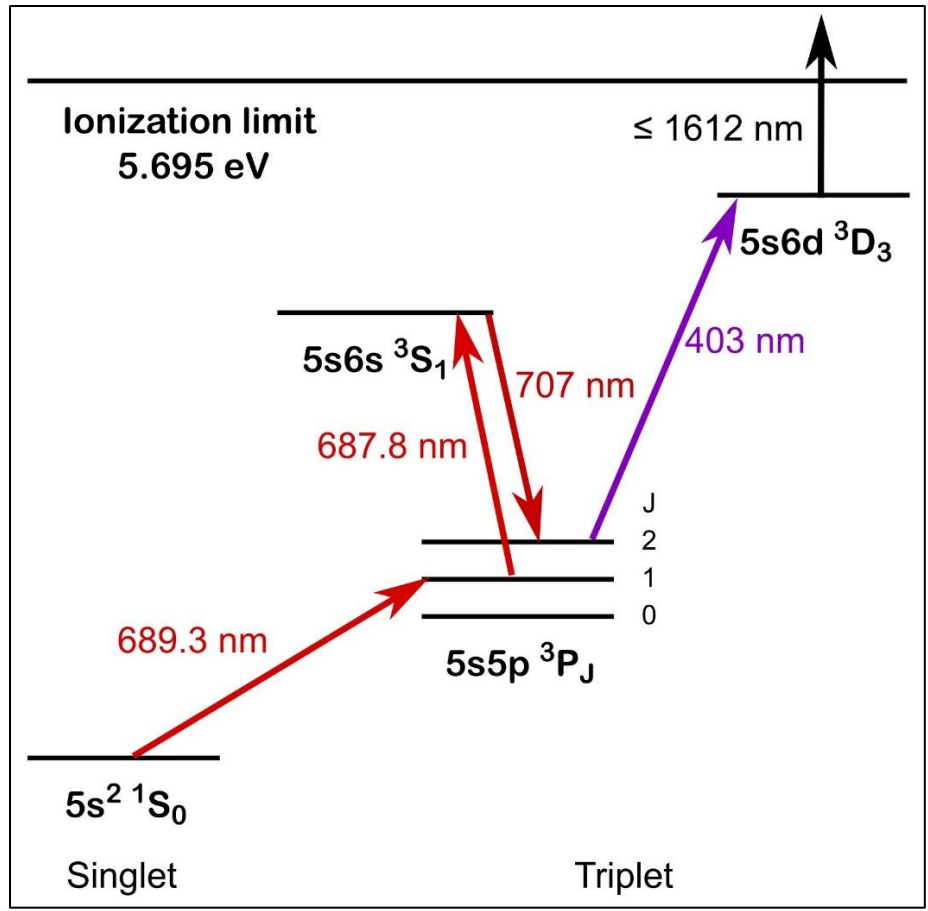

**Fig. 4.** Grotrian diagram for atomic strontium, showing the laser transitions needed for ionization (not to scale).

**Applications of MRLIS**

In this section we discuss the application of MRLIS to atomic strontium (Sr). We estimate that the combination of high intra-cavity power and long transit time leads to near unity in ionization probability. The quoted spectroscopic data is available online [40]. We analyze here the case of strontium (Sr), which has a rare isotope, Sr-84, that can be used as a tracer for bone absorption to detect the onset of osteoporosis [41]. This same isotope can also be used to produce the radioisotope Sr-85 which has been proposed for a new test of quantum mechanics [42].

A schematic of the relevant energy levels in atomic strontium is shown in Fig. 4. The ionization limit of Sr is an energy of 5.695 eV. The ground $^1S_0$ state would first be excited by a laser near 689.3 nm to the $^3P_1$ state which has a decay rate of $4.69 \times 10^4$ s$^{-1}$. A second laser near 687.8 nm would then excite atoms from the $^3P_1$ state to the $^3S_1$ state which has a decay rate of $2.7 \times 10^7$ s$^{-1}$. Atoms will scatter the laser at 687.8 nm until the upper state decays to a long-lived metastable state $^3P_2$, while emitting a photon at 707 nm. Approximately 80% of the atoms will be optically pumped into this metastable state, which can be improved to 100% by adding a repump laser from the $^3P_0$ state at 679.1 nm.

After preparation of atoms in the $^3P_2$ state, they would propagate to the optical resonator and would be excited by a laser near 403 nm to the $^3D_3$ state which has a decay rate of approximately $2 \times 10^7$ s$^{-1}$. While the atoms are cycling on this transition at 403 nm, they would be ionized in the cavity by a laser tuned to a wavelength that is shorter than 1612 nm. The laser will be scanned just above the ionization limit to optimize ionization with an autoionizing state, but it may be simpler to find a convenient high-power laser that is above the ionization threshold. Positive ions will be accelerated by an electric field that will remove the desired isotopes from the beam and collect them on a cathode plate.

**Discussion**

Our proposed method for efficient isotope separation based on laser ionization can be compared with the prior method of laser ionization by simple scaling. In that case, pulsed lasers were utilized, with a typical pulse duration of 100 ns and a pulse repetition rate of 10 kHz [9]. The average laser power was in the range of 20 Watt as typical with pulsed dye lasers. The enhancement of the ionizing laser in MRLIS of over 1000x and use of continuous-wave solid state lasers with a power of 5-10 W would provide a comparable laser intensity to the pulsed laser, assuming similar focusing. The interaction time would be about 1000x longer in the case of MRLIS, addressing all the desired isotopes in the atomic beam, and bringing the ionization probability to near unity. In addition, the use of narrow-band continuous-wave lasers would enable very high isotopic purity, unlike the previous method where the excitation lasers had large linewidths on the order of 100 MHz, limiting isotopic purity.

The most exciting application of MRLIS is towards production of radioisotopes as an alternative to radiochemistry, due to the small footprint of the system which could fit into standard hot cells. The most compelling case is production of the radioisotope Lu-177 from stable Lu-176 due to the much larger (~700x) neutron cross section of Lu-176 compared with Yb-176 [43]. The latter stable isotope is currently used to produce Lu-177 which is separated using radiochemistry [44]. There are many other pairs of stable/radioactive isotopes of the same element that have important medical applications, most notably Mo-98/Mo-99 [45].

In conclusion, isotopes are an important and valuable resource for humanity, and improved methods for their separation will enable new applications, especially important for medicine. We expect that MRLIS will be an important step forward in that direction.

We thank Ajith Jothikumari for help with figures and for comments on the manuscript. We also thank Logan Hillberry and Henry Chance for comments on the manuscript and acknowledge support from the Sid W. Richardson Foundation and the Novo Nordisk Foundation.

**Appendix**

Cavity transmission for a Gaussian beam propagating in a Fabry-Perot etalon was numerically modeled using open-source MATLAB code [29]. The cavity mirror spacing was set at 50 mm, with incident beam spot sizes modeled in a range from 20 mm to 1 mm. Results were computed for both 99% and 99.9% etalon mirror reflectivity. Results for the propagation of incident plane waves in the etalon were also calculated.

For a given etalon, transmission as a function of wavelength is characterized by its interferometer transfer function (ITF):

$$ITF(\lambda) = \frac{I_{\text{out}}(\lambda)}{I_{\text{in}}(\lambda)}$$

where $I_{in}$ is the intensity of light incident on the etalon, and $I_{\text{out}}$ is the intensity of light exiting the etalon. The outgoing intensity can be found by the spatial integration of the outgoing field:

$$I_{\text{out}} = \int_0^\infty |U_{\text{out}}(r,z)|^2 2\pi r \, dr$$

where $r$ is the radial axis and $z$ is the optical axis, in the direction of the incoming beam. The total outgoing field is the sum of contributions by exiting beams that have made $m$ round trips between the cavity mirrors, where $m = 0,1,2,...$ For incident plane waves, these partial beams are expressible as a geometric series, and their sum by an Airy function.

Incident Gaussian beams require a more general approach. Here, we use the method of ray transfer matrix analysis, which allows the performance of ray tracing calculations in cases where the paraxial approximation is valid. In this method, a $2 \times 2$ ray transfer matrix (sometimes called an ABCD matrix), operates on a vector representing an incoming light ray, yielding a new vector representing the outgoing ray:

$$\begin{bmatrix} q_{out} \\ 1 \end{bmatrix} = k \begin{bmatrix} A & B \\ C & D \end{bmatrix} \begin{bmatrix} q_{in} \\ 1 \end{bmatrix}$$

Here, $k$ is a normalization constant, and $q$ is the complex beam parameter defined by:

$$\frac{1}{q} = \frac{1}{R} - i \frac{\lambda}{\pi n w^2}$$

where $R$ is the Gaussian beam's radius of curvature, $w$ is its spot size, $n$ is the index of refraction of the material through which the beam is passing, and $\lambda$ is its wavelength. Because ABCD matrices are linear operators, matrices representing each optical element in a system can be multiplied together to give a single resultant matrix representing the system as a whole:

$$M = M_N M_{N-1} \cdots M_2 M_1$$

Modeling a vacuum spaced etalon requires three different kinds of ABCD matrix. First is $M_{refr}$, representing the refraction of a ray either entering or exiting the etalon:

$$M_{refr} = \begin{bmatrix} 1 & 0 \\ 0 & n_1/n_2 \end{bmatrix}$$

where $n_1$ is the index of refraction in the initial medium and $n_2$ is the index of refraction in the final medium. The second ray transfer matrix is $M_{prop}$, which represents the propagation of rays through a homogenous medium:

$$M_{prop} = \begin{bmatrix} 1 & h/n \\ 0 & 1 \end{bmatrix}$$

here $h$ is the distance propagated (in our case, the mirror-to-mirror distance), and $n$ is the index of refraction of the medium propagated through (vacuum, so here $n = 1$). Third and last is $M_{refl}$, which gives the reflection of rays from a planar surface:

$$M_{refl} = \begin{bmatrix} 1 & 0 \\ 0 & 1 \end{bmatrix}$$

Conceptually, it is natural to model the propagation of the beam through the etalon in three parts: $M_{in}$, in which a ray enters the etalon through the first mirror, $M_{cav}$, in which the ray bounces back and forth between the cavity mirrors, and finally $M_{out}$, in which the ray exits the etalon. $M_{in}$ only involves refraction by the first mirror:

$$M_{in} = M_{refr}$$

For $M_{cav}$, consider the constituent matrices of a round trip through the interior of the etalon: 1) propagation through the interior to the surface of the second mirror, 2) reflection from the second mirror's surface, 3) propagation back to the first mirror, and 4) reflection from the first mirror's surface. In matrix form:

$$M_{cav} = M_{refl}^{\leftarrow} M_{prop}^{\leftarrow} M_{refl} M_{prop}$$

$M_{out}$, in which the ray is transmitted through the cavity, is comparatively straightforward: propagation to the second mirror surface, followed by refraction through the second mirror to exit the etalon:

$$M_{out} = M_{refr} M_{prop}$$

Now, $M_{in}$ and $M_{out}$ occur exactly once for each transmitted ray, but the number $m$ of round trips within the cavity can be any non-negative integer, so the overall system matrix for any possible transmitted ray has the form:

$$M = M_{out}(M_{cav})^m M_{in}$$

From this system matrix, the transmitted field can be computed from the relation:

$$U_{out}(r, z; M) = \frac{1}{A + B/q(z_{in})} \exp(-ik \frac{r^2}{2q(z_{out})})$$

where $z_{in}$ and $z_{out}$ are the axial positions at the entrance and exit of the etalon, respectively.

The ABCD ray-tracing model does not capture changes in field amplitude, so that must be included in the final summation of contributions to the field from all transmitted beams:

$$U_{out}(r, z) = \sum_{m=0}^{\infty} A_m U_m, \quad A_m = t_1 t_2 (r_1 r_2)^m$$

Here, $t_1, t_2$ and $r_1, r_2$ are the transmittivity and reflectivity coefficients of the first and second mirrors. The above provides a method to compute the ITF for any chosen wavelength, given the incoming beam spot and waist size, the indices of refraction for intra- and extracavity material, the mirror-to-mirror spacing, and the reflectivity of each mirror.

While the number of cavity round trips $m$ used in computing the transmitted field ranges from zero to infinity, because the partial field contribution becomes smaller and smaller for higher $m$ terms, the number of computed round trips can be truncated at a finite value. This truncation was chosen by considering the output ITF values to have converged once the most recent term changes the overall transmitted field amplitude by <0.001%, typically $m \sim 100$. To capture fine details, $10^3$ points/nm were computed in the vicinity of the transmission peaks, with $10^2$ points/nm in the inter-peak region. The plane wave computations were performed by setting the spot size to several orders of magnitude larger than the cavity dimensions.

**References**

[1] https://world-nuclear.org/information-library/non-power-nuclear-applications/radioisotopes-research/radioisotopes-in-medicine.aspx

[2] "Brachytherapy: a comprehensive review," Y.K. Lim and D. Kim, Medical Physics **32**, 25 (2021).

[3] "Targeted radionuclide therapy: a historical and personal review," S.J. Goldsmith, Seminars in Nuclear Medicine *50*, 87 (2020).

[4] "Radiopharmaceuticals for PET and SPECT imaging: a literature review over the last decade," G. Crisan et al., Int. J. Mol. Sci. **23**, 5023 (2022).

[5] "Applications of stable, nonradioactive isotope tracers in in-vivo human metabolic research," I-Y Kim et al., Experimental and Molecular Medicine, **48**, 203 (2016).

[6] "Electromagnetic separation of isotopes at Oak Ridge," L. O. Love, Science **182**, 343 (1973).

[7] "The separation of isotopes by centrifuging," J. W. Beams and F. B. Haynes, Phys. Rev. **50**, 491 (1936).

[8] https://www.osti.gov/biblio/1298983

[9] https://www.ornl.gov/news/generations-separation-emis-keeps-improving

[10] P. A. Bokhan *et al.*, *Laser Isotope Separation in Atomic Vapor* (Wiley-VCH Verlag GmbH & Co. KGaA, Weinheim, 2006).

[11] "Stable isotope production of 168Yb and 176Yb for industrial and medical applications," H. Park et al., J. Nucl. Sci. and Tech. Suppl. **6**, 111 (2008).

[12] "Method for isotope separation of thallium," D-Y Jeong et al., US Patent 7,323,651 B2, Jan. 29, 2008.

[13] "Numerical study of the laser isotope separation of optically pumped 102Pd," M.V. Suryanarayana and M. Sankari, Scientific Reports 14, 4080 (2024)

[14] "Laser isotope separation of 203Tl through pulsed laser optical pumping," M.V. Suryanarayana, J. Opt. Soc. B **41**, 311 (2024).

[15] "Semiconductor lasers: fundamentals and applications," A. Baranov and E. Tournié, eds. 2013.

[16] "Optically pumped VECSELs: review of technology and progress," M. Guina, A. Rantamäki, and A. Härkönen, J. Phys. D. **50**, 383001 (2017).

[17] "Efficient isotope separation with single-photon atomic sorting," M. Jerkins, I. Chavez, U. Even, and M. G. Raizen, Phys. Rev. A **82**, 033414 (2010).

[18] "Magnetically activated and guided isotope separation," M. G. Raizen, and B. G. Klappauf, New J. Phys. **14**, 023059 (2012).

[19] "Demonstration of magnetically activated and guided isotope separation," T.R. Mazur, B. Klappauf, and M.G. Raizen, Nature Physics **10**, 601 (2014).

[20] Raizen, M.G., Klappauf, B.G., Isotope separation by magnetic activation and separation. United States Patent #8,672,138.

[21] "Lutetium-177 Therapeutic Radiopharmaceuticals: Linking Chemistry, Radiochemistry, and Practical Applications", S. Banerjee, M.R. Pillai, and F.F. Knapp., Chemical Reviews **115**, 2934 (2015).

[22] "Laser cooling of atoms," H. Metcalf and P. van der Straten, 1999 (Berlin: Springer).

[23] "Brightening of a supersonic beam of neutral atoms," E. Anciaux, G. Stratis, and M.G. Raizen, Phys. Scr. **93**, 124009 (2018).

[24] "Nuclear and radiochemistry: Fundamentals and applications," J.-V. Kratz, John Wiley & Sons, 2022.

[25] "Stimulated Raman adiabatic passage in physics, chemistry, and beyond," N.V. Vitanov, A.A. Rangelov, B.W. Shore, and K. Bergmann, Reviews of Modern Physics **89**, 015006 (2017).

[26] "Lasers" A. E. Siegman, University Sciences Books, 1986.

[27] "A marginally stable optical resonator for enhanced atom interferometry," I. Riou et al., J. Phys. B **50**, 155002 (2017).

[28] "Manufacture of a 10-km-scale radius-of-curvature surface by use of a thin-film coating technique," S. Miyoki et al., Opt. Lett. **30**, 1399 (2005).

[29] "ABCD transfer matrix model of Gaussian beam propagation in Fabry-Perot etalons," D. Martin-Sanchez et al., Opt. Express **30**, 46404 (2022).

[30] David Reitze, private communication.

[31] "AC Stark shift of atomic energy levels," N.B. Delone and V.P. Krainov, Physics-Uspekhi **42**, 669 (1999).

[32] "Physics of optical lattice clocks," A. Derevianko and H. Katori, Rev. Mod. Phys. **83**, 331 (2011).

[33] "Far-off-resonance optical trapping of atoms," J.D. Miller, R.A. Cline, and D.J. Heinzen, Phys. Rev. A **47**, R4567 (1993).

[34] "Using stable isotopes to assess mineral absorption and utilization by children," S.A. Abrams, Amer. Jour. of Clinical Nutrition **70**, 955 (1999).

[35] K. Schilling et al., "Urine metallomics signature as an indicator of pancreatic cancer," K. Schilling et al., Metallomics **12**, 752 (2020).

[36] "Frequency modulation (FM) spectroscopy: theory of lineshapes and signal-to-noise analysis, G.C. Bjorklund, M.D. Levenson, W. Lenth, and C. Ortiz, Applied Physics B **32**, 145 (1983).

[37] "Spectroscopy with squeezed light," E.S.Polzik, J.Carri, and H.J.Kimble, Phys. Rev. Lett. **68**, 3020 (1992).

[38] "Quantum projection noise: population fluctuations in two-level systems," W.M. Itano et al., Phys. Rev. A **47**, 3554 (1993).

[39] "Ultrasensitive isotope trace analysis with a magneto-optical trap," C.Y. Chen et al., Science **286**, 542 (1999).

[40] https://physics.nist.gov/PhysRefData/ASD/lines_form.html

[41] "Isotopic biomarkers for rapid assessment of bone mineral balance in biomedical applications," A. Anbar et al., Patent WO2011156583A1.

[42] "Do atoms age?" M.G. Raizen, D.E. Kaplan, S. Rajendran, Phys. Lett. B **832**, 137224 (2022).

[43] "Production of 177Lu for targeted radionuclide therapy," A. Dash, M. Pillai, F. Knapp, Nuclear Medicine and Molecular Imaging **49**, 85 (2015).

[44] "Nuclear and radiochemistry: fundamentals and applications," J.V. Kratz, John Wiley & Sons, 2022.

[45] "Molybdenum-99 for medical imaging", National Academy of Sciences, 2016.